\documentclass[pre,aps,showpacs,superscriptaddress,twocolumn,floatfix,nofootinbib]{revtex4-1}

\usepackage{dsfont}
\usepackage{mathtext}
\usepackage[cp1251]{inputenc}

\usepackage{mathtools}
\usepackage[usenames,dvipsnames]{xcolor}
\usepackage{amsmath}
\usepackage{amssymb,mathrsfs}
\usepackage{graphicx}
\usepackage{epstopdf}
\usepackage[colorlinks=true]{hyperref}

\usepackage{mathtext}

\usepackage{etoolbox} 
\usepackage{lipsum} 
\usepackage[capitalize]{cleveref}

\makeatletter
\appto{\appendix}{%
	\@ifstar{\def\theequation@prefix{A.}}%
	{}%
}
\makeatother

\newcommand{\al}{\alpha}
\newcommand{\om}{\omega}

\newcommand{\tk}{\widetilde{k}}
\newcommand{\tom}{\widetilde{\omega}}
\newcommand{\tal}{\widetilde{\alpha}}

\newcommand{\prt}{\partial}

\newcommand {\ox} {\overline{x}}

\newcommand{\rom}[1]{\uppercase\expandafter{\romannumeral #1\relax}}

\graphicspath{{Figures/}}

\begin{document}

\title{Formation of dispersive shock waves in evolution of a two-temperature collisionless plasma}

\author{Sergey~K.~Ivanov}
\affiliation{Moscow Institute of Physics and Technology, Institutsky lane 9, Dolgoprudny, Moscow region, 141700, Russia}
\affiliation{Institute of Spectroscopy, Russian Academy of Sciences, Troitsk, Moscow, 108840, Russia}

\author{Anatoly~M.~Kamchatnov}
\affiliation{Moscow Institute of Physics and Technology, Institutsky lane 9, Dolgoprudny, Moscow region, 141700, Russia}
\affiliation{Institute of Spectroscopy, Russian Academy of Sciences, Troitsk, Moscow, 108840, Russia}

\begin{abstract}
The nonlinear dynamics of pulses in a two-temperature collisionless plasma with formation 
of dispersion shock waves is studied. An analytical description is given for arbitrary form of an initial 
disturbance with smooth enough density profile on a uniform density background. For large time
after the wave breaking moment dispersive shock waves are formed. Motion of their edges is studied 
in framework of Gurevich-Pitaevskii theory and Whitham theory of modulations. The analytical results are 
compared with numerical solution.
\end{abstract}




\maketitle

\section{Introduction} \label{sec1}

The study of nonlinear waves is one of fascinating topics in modern science and
it has attracted much attention in many various fields of research.
In particular, it is now well known that a typical evolution of an initial pulse with a fairly
smooth and large initial profile is accompanied by a gradual steepening followed by the wave
breaking and formation of dispersive shock wave (DSW). The initial stage of evolution
admits a purely hydrodynamic description where the classical Riemann method seems quite
convenient (see, e.g., \cite{FRW-2009,IvanovKamchatnov-2019,IKP-2019,IKP-2019F,IKP-2020}).
After the wave breaking moment the oscillatory wave structures emerge in the
evolution of pulses and they are called ``dispersive shock waves'' or ``undular bores''.
Nowadays DSWs are recognized as a universal physical phenomenon encountered in diverse
areas of science (see, e.g., \cite{kamch-2000,eh-2016}). The notion of DSWs was introduced
theoretically and demonstrated experimentally in water waves
\cite{Bazin-65,DarcyBazin-65,BenjaminLighthill-54},
atmospheric layers \cite{SmythHolloway-88,Christie-92,HD-12},
Bose-Einstein condensates \cite{DBSH-01,secscb-05,KGK-04,hacces-06},
waves in magnetics \cite{magn-17},
intense electron beams \cite{MKFHBOT-13},
in nonlinear optics \cite{RG-89,WJF-07,Kodama-99},
and, most notably, in plasma
\cite{ABS-68,TBI-70,GM-84,MS-63,Karpman-74} dynamics
where DSWs were observed experimentally for the first time in the laboratory set up.
Plasma shock waves have been studied in different plasma environments,
such as ion-acoustic \cite{TBI-70,SN-05,Romagnani-08,BNS-08}, dust-ion-acoustic
\cite{NBS-99,Nakamura-02,PGL-01}
and in dusty multi-ion \cite{MCS-09,Duha-09} plasma.
In this paper, we focus on theoretical description of ion-acoustic DSWs in a two-temperature
collisionless plasma.

The simplest and, apparently, the most powerful theoretical approach to description of DSWs was formulated
by Gurevich and Pitaevskii \cite{gp-73} in framework of the Whitham theory of modulation of nonlinear waves
\cite{Whitham-65,Whitham-74} which was based on large difference between scales of the wavelength
of nonlinear oscillations within DSW and the size of the whole DSW. 
In plasma physics applications, the dynamics of waves can be described by different wave equations, viz.,
Korteweg-de Vries (KdV) equation \cite{WashimiTaniuti-74}, KdV-Burgers (KdVB) equation \cite{MS-63,Sagdeev-66},
modified KdV (mKdV) \cite{ChanteurRaadu-87}, Gardner equation \cite{RTP-08},
nonlinear Schr\"odinger (NLS) equation \cite{Zakharov-72}, derivative NLS (DNLS) equation 
\cite{Rogister-71,Mjolhus-76}. For most of these equations (except for the KdVB equation) 
the corresponding modulation system can be reduced
to a diagonal (Riemann) form. It is known that the possibility of diagonalization of the system of Whitham
equations is closely related with their full integrability. However, although a large number of equations of
nonlinear plasma physics are of this type, there are important situations when it is not the case and
one has to resort to equations that are not completely integrable. A typical example of such a
situation is presented by the nonlinear ion-sound waves (see, e.g., \cite{GM-84,gke-90-plasma}),
and in these cases some development of the general theory of DSWs is required.

Apparently, the first general statement in this area was devised by Gurevich and Meshcherkin \cite{GM-84},
who proposed the condition which replaces the ``jump condition'' known in the theory viscous shocks. This condition is
formulated in terms of Riemann invariants of the hydrodynamic system obtained in the dispersionless approximation of
the original nonlinear wave equations under consideration. Typically, wave breaking occurs in the simple-wave
flow when all physical variables of the system can be expressed as functions of only one of them what means
that after transformation to the corresponding Riemann invariants only one of them breaks and the others
remain constant. Gurevich and Meshcherkin claimed that this statement is correct also after formation of the DSW,
so that flows at both its edges have the same values of the non-breaking Riemann invariants. This property is
evidently correct in a simple case of self-similar solutions of Whitham equations for the KdV equation studied in
\cite{gp-73}, when Whitham equations are transformed to the diagonal form, and Gurevich and Meshcherkin generalized
it to situations when Riemann invariants of the modulation equations are unknown or even do not exist.

The next important success in this direction was achieved by G.~El in Ref.~\cite{El-05}, where
it was noticed that the Whitham modulation equations degenerate at the DSW edges, where they match with the
simple-wave dispersionless solution, into ordinary differential equations whose solutions provide
the dependence of the DSW characteristics at the corresponding edge on a single parameter.
This method made it possible to find the main parameters of DSWs arising due to evolution
of the initial step-like discontinuities in many non-integrable situations including the 
case of ion-sound waves \cite{El-05}. Recently
it was shown in Ref.~\cite{Kamchatnov-19} that El’s method can be substantially generalized to arbitrary
initial simple-wave conditions of the Gurevich-Meshcherkin class.
When such a simple wave breaks, then, according to El~\cite{El-05}, we know the limiting expressions
for the characteristic velocities of the Whitham system at the DSW edges from simple
physical considerations --- they are equal either to the group velocity of the wave at
the small-amplitude edge, or the soliton velocity at the soliton edge. This
information, together with the known smooth solution of the dispersionless limit,
is enough for finding the laws of motion of the corresponding edges of the DSW for an
arbitrary initial profile of the simple-wave type. This approach was successfully used
in \cite{IvanovKamchatnov-19} for theoretical description of DSWs in fully nonlinear
theory of shallow water waves described by the Serre equation, and we will use here this
method for finding the laws of motion of DSW edges in ion-acoustic DSWs.

The method of Refs.~\cite{El-05,Kamchatnov-19} is based on the universal applicability
of Whitham’s  ‘number of waves conservation law’
\cite{Whitham-65,Whitham-74}
\begin{equation}\label{eq1}
\frac{\prt k}{\prt t}+\frac{\prt \om(k)}{\prt x}=0,
\end{equation}
Here $k=2\pi/L$ and $\om=kV$ are the wave vector and the frequency
of a linear harmonic wave $\propto \exp[i(kx - \om t)]$ propagating along a smooth
background, $L$ is the wavelength of the nonlinear periodic wave,
$V$ being the phase velocity of the periodic wave. Long ago, Stokes noticed \cite{stokes}
(see also \cite{lamb}) that the solitons's velocity can be found from the linear
dispersion law because the soliton's tails obey the same linearized equations and
propagate with the same velocity as the whole soliton. This yields the expression
for the soliton velocity $V_s={\tom(\tk)}/{\tk}$, where
$\tom(\tk)\equiv-i\om(i\tk)$, $\tk$ being the inverse half-width of the soliton.
It is natural to suppose \cite{El-05,Kamchatnov-19} that for the simple-wave type
of initial conditions the soliton counterpart of Eq.~(\ref{eq1}) is valid
\begin{equation}\label{eq2}
\frac{\prt \tk}{\prt t}+\frac{\prt \tom(\tk)}{\prt x}=0.
\end{equation}
El showed in Ref.~\cite{El-05} that Eqs.~(\ref{eq1}) and (\ref{eq2}) reduce at the
corresponding edges of the DSW to the ordinary differential equations and
their solutions gives at once the edges velocities of the DSW for the case of
Gurevich-Pitaevskii initial step-like problem. This problem for the
ion-acoustic case was solved in Refs.~\cite{ekt-03,El-05}.  In this paper,
we generalize this theory to arbitrary form of the initial pulse
and apply the method of Ref.~\cite{Kamchatnov-19} to investigation of
evolution of initial simple-wave ion-sound pulses.

The paper is structured as follows: the basic equations describing the dynamics of
ion-acoustic waves in plasma are written out in Section~\ref{sec2}. The
problem of the expansion of the initial hump of an ion density is considered in
Section~\ref{sec3} and solved in the dispersionless approximation by the Riemann method.
This solution can be applied to evolution of the pulse before the wave breaking moment.
As is known, in the general case the initial pulse splits eventually to two pulses
propagating in opposite directions, and each such a pulse can be represented as a
simple wave so the method of Ref.~\cite{Kamchatnov-19} becomes applicable.
In Section~\ref{sec4}, we find by this method the main characteristics of the DSW
arising after breaking of a simple-wave.

\section{Main equations} \label{sec2}

We consider the system of equations which describe finite-amplitude ion-acoustic waves
in a two-temperature ($T_e \gg T_i$) collisionless plasma
\cite{Karpman-74,Sagdeev-1966,gke-90-plasma}
\begin{equation} \label{PlasmaDimention}
\begin{split}
	& n_T+(nv)_X = 0, \\
	& v_T + v v_X + \frac{e}{m_i} \phi_X = 0, \\
	& \phi_{XX} = 4 \pi e \left( n_0 \exp{\frac{e\phi}{T_e}} - n \right),
\end{split}
\end{equation}
where $n$ is the ion density, $v$ is the hydrodynamic velocity of
the ions, $m_i$ is their mass, $\phi$ is the electric potential, $e$ and $T_e$
are the charge and temperature of electrons, and $n_0$ is the
density of an undisturbed plasma. By means of replacements
\begin{equation} \label{}
\begin{split}
	t & = \Omega T, \quad x = D^{-1} X, \\
	\rho & = \frac{n}{n_0}, \quad u = \frac{v}{c_s}, \quad \varphi = \frac{e}{T_e} \phi,
\end{split}
\end{equation}
where $\Omega = (4\pi e^2n_0/m_i)^{1/2}$ is the ion plasma frequency,
$D = (T_e/4\pi e^2n_0)^{1/2}$ is the Debye length and $c_s^2=T_e/m_i$ is sound velocity,
Eqs.~(\ref{PlasmaDimention}) are transformed to a dimensionless form
\begin{equation} \label{PlasmaEq}
\begin{split}
& \rho_t+(\rho u)_x = 0, \\
& u_t + u u_x +  \varphi_x = 0, \\
& \varphi_{xx} =  e^{\varphi} - \rho.
\end{split}
\end{equation}
This system supports periodic traveling waves which have linear and solitary wave limits and
also possesses at least four conservation laws \cite{gke-90-plasma}.
The harmonic dispersion law follows from linearized Eqs.~(\ref{PlasmaEq})
\begin{eqnarray} \label{harmlaw}
\om(k,\rho,u) = k\left( u + \frac{1}{\sqrt{1+k^2/\rho}} \right),
\end{eqnarray}
so that
\begin{eqnarray} \label{sollaw}
\tom(\tk,\rho,u) = \tk\left( u + \frac{1}{\sqrt{1-\tk^2/\rho}} \right).
\end{eqnarray}

In smooth flows, when the terms with higher space derivatives can be neglected, we arrive
at the dispersionless limit with $\varphi= \ln{\rho}$, which yields the Euler isothermal 
gas-dynamic equations with the equation of state $p(\rho) = \rho$:
\begin{equation} \label{hydro}
\begin{split}
& \rho_t+(\rho u)_x = 0, \\
& u_t + u u_x +  \frac{\rho_x}{\rho} = 0.
\end{split}
\end{equation}
The solution of such equations is greatly simplified if we pass from the
ordinary physical variables $\rho$, $u$ to the so-called
``Riemann invariants'' (see, e.g., \cite{LandauLifshitz-1987}). For Eqs.~(\ref{hydro})
the Riemann invariants are well known and can be written as
\begin{eqnarray} \label{riminv}
r_{\pm} = u \pm \ln {\rho}.
\end{eqnarray}
In these variables the hydrodynamic equations (\ref{hydro}) take simple symmetric form
\begin{equation} \label{rimeq}
\frac{\prt r_{\pm}}{\prt t}+v_{\pm}(r_{-},r_{+})\frac{\prt r_{\pm}}{\prt x}=0,
\end{equation}
where
\begin{equation} \label{rimvel}
v_\pm = u \pm 1
\end{equation}
are the characteristic velocities of the system (\ref{hydro}). They are expressed
in terms of the Riemann invariants by the relations
\begin{equation} \label{rimvel2}
v_\pm = \frac{1}{2}(r_++r_-)\pm1.
\end{equation}
These velocities coincide exactly with the corresponding
limits of the Whitham velocities what guarantees a
continuous matching of the dispersionless flows with the DSW at both its edges.

Having formulated the problem, we proceed to studying the dispersionless evolution of
the plasma density pulse.

\section{Dispersionless solution} \label{sec3}

We assume that the initial state of plasma can be represented as a density hump
on the constant uniform background and the density profile can be considered fairly
smooth at the initial stage of evolution, so we turn to the system of hydrodynamic
equations (\ref{hydro}) or (\ref{rimeq}). Riemann noticed that Eqs.~(\ref{rimeq}) can be
linearized by the hodograph transform (see, e.g., Refs. \cite{kamch-2000,LandauLifshitz-1987})
in which one considers $x$ and $t$ as functions of the independent variables
$r_\pm$. This results in the following system of linear equations:
\begin{equation}\label{HodTransEq}
\begin{split}
\frac{\partial x}{\partial r_-}-v_{+}(r_{-},r_{+})\frac{\partial t}{\partial r_-} & = 0, \\
\frac{\partial x}{\partial r_+}-v_{-}(r_{-},r_{+})\frac{\partial t}{\partial r_+} & = 0.
\end{split}
\end{equation}
It should be noticed that the Jacobian of this transformation is equal to
\begin{equation}\label{}
\begin{split}
J=\left| \frac{\partial (x,t)}{\partial (r_+,r_-)} \right| =
\frac{\partial t}{\partial r_+}\frac{\partial t}{\partial r_-} (v_--v_+),
\end{split}
\end{equation}
and hence the hodograph transform breaks down whenever
${\partial t}/{\partial r_+}=0$ or ${\partial t}/{\partial r_-}=0$.
This means that the hodograph transform is applicable only to the general solution,
when both Riemann invariants are varying during the wave evolution.
We look for the solution of the system (\ref{HodTransEq}) in the form
\begin{equation}\label{HodTransEqSol}
\begin{split}
x-v_{+}(r_{-},r_{+})t & = w_+(r_-,r_+), \\
x-v_{-}(r_{-},r_{+})t & = w_-(r_-,r_+),
\end{split}
\end{equation}
where the functions $w_{\pm}$ are to be found. Then we get
\begin{equation}\label{tEq}
\begin{split}
t=-\frac{w_+-w_-}{v_+-v_-}.
\end{split}
\end{equation}
Differentiation of Eqs.~(\ref{HodTransEqSol}) with respect to $r_{\pm}$ gives the
relations $-\prt v_{\pm}/\prt r_{\mp}t= \prt w_{\pm}/\prt r_{\mp}$ and
elimination of $t$ by means of (\ref{tEq}) shows that the unknown functions
$w_\pm(r_+, r_-)$ should satisfy the Tsarev equations \cite{Tsarev-1991}
\begin{equation}\label{Tsarev}
\begin{split}
\frac1{w_+-w_-}\frac{\prt w_+}{\prt r_-}=
\frac1{v_+-v_-}\frac{\prt v_+}{\prt r_-}, \\
\frac1{w_+-w_-}\frac{\prt w_-}{\prt r_+}=
\frac1{v_+-v_-}\frac{\prt v_-}{\prt r_+}.
\end{split}
\end{equation}
Now we notice that since the velocities $v_\pm$ are given by
expressions (\ref{rimvel2}), the right-hand sides of both Eqs.~(\ref{Tsarev})
are equal to each other:
\begin{equation}\label{58}
\frac{1}{v_+-v_-}\frac{\partial v_+}{\partial r_-}=\frac{1}{v_+-v_-}\frac{\partial v_-}{\partial r_+}.
\end{equation}
Consequently $\partial w_+/ \partial r_- = \partial w_-/ \partial r_+$
and $w_\pm$ can be sought in the form
\begin{equation}\label{58-2}
w_+=\frac{\prt W}{\prt r_+},\quad w_-=\frac{\prt W}{\prt r_-}.
\end{equation}
Substitution of Eqs.~(\ref{58}) and (\ref{58-2}) into Eqs.~(\ref{Tsarev}) shows
that the function $W$ obeys the Euler-Poisson equation
\begin{equation}\label{58-3}
\frac{\prt^2W}{\prt r_+\prt r_-}-\frac{1}{4}\left(\frac{\prt W}{\prt r_+}-\frac{\prt W}{\prt r_-}\right)=0
\end{equation}
A formal solution of Eq.~(\ref{58-3}) in the $(r_+, r_-)$ plane (the
so-called hodograph plane) can be obtained with the use
of the Riemann method (see, e.g., \cite{Sommerfeld-49,CourantHilbert-62}).

In his fundamental paper \cite{Riemann-1860}, B.~Riemann gave
the following method of solving this
problem. If we are interested in the value of the function
$W$ at the point $P = (\xi,\eta)$ in the hodograph plane $(r_+,r_-)$, then we should
draw in it from the point $P$ two characteristics ($r_+=\xi = \mathrm{const}$)
and ($r_-=\eta = \mathrm{const}$), which together with the
boundaries with known values of $W(r_+, 0)$ and $W(0,r_-)$
along them form a closed contour $\mathcal{C}$ in this plane.
The symbols $(\xi,\eta)$ denote here the coordinates of the
``observation point'' in the hodograph plane, whereas
the notation $(r_+,r_-)$ is used for varying along the contour
$\mathcal{C}$ coordinates.
Riemann showed that $W(P)$ can be represented in the form
\begin{equation}\label{m1-282.9}
W(P)=\frac12(R\overline{W})_A+\frac12(R\overline{W})_B\pm\int_A^B(Vdr_++Udr_-),
\end{equation}
where the points $A$ and $B$ are projections of the ``observation'' point $P$
to $\mathcal{C}$ along the $r_+$ and $r_-$ axis respectively. Here
\begin{equation}\label{m1-284.4}
\begin{split}
& U=\frac12 \left(R\frac{\prt \overline{W}}{\prt r_-}-
\overline{W}\frac{\prt R}{\prt r_-}\right)-\frac14\overline{W}R,\\
& V=\frac12 \left(\overline{W}\frac{\prt R}{\prt r_+}
-R\frac{\prt \overline{W}}{\prt r_+}\right)-\frac14\overline{W}R,
\end{split}
\end{equation}
where $\overline{W}$ denotes such a contraction of $W$
on the $\mathcal{C}$ curve; the functions $\overline{W}$ are known from the
initial conditions for the problem under consideration.
The function $R$ is called the Riemann function and it satisfies the equation
\begin{equation}\label{eq16b}
\frac{\prt^2R}{\prt r_+\prt r_-}+\frac14\left(\frac{\prt R}{\prt r_+}-
\frac{\prt R}{\prt r_-}\right)=0.
\end{equation}
Besides that, Riemann imposed on it the following additional conditions:
\begin{equation}\label{BoundR}
\begin{split}
\frac{\partial R}{\partial r_+} -\frac14 R = & 0 \quad \text{along the characteristic} \quad r_-=\eta, \\
\frac{\partial R}{\partial r_-} +\frac14 R = & 0 \quad \text{along the characteristic} \quad r_+=\xi, \\
\end{split}
\end{equation}
and $R(\xi,\eta;\xi,\eta)=1$.
These expressions suggest that $R$ can be looked for in the form
\begin{equation}\label{}
\begin{split}
R = \exp{\left[ \frac14 (r_+-\xi-r_-+\eta) \right]}  F\left(r_+,r_-;\xi,\eta\right).
\end{split}
\end{equation}
Substituting it into Eq.~(\ref{eq16b}),
we obtain the Bessel equation for the function $F$,
which shows that $R$ can be written in the form
\begin{equation}\label{Rim-func}
\begin{split}
\begin{split}
R =& \exp{\left[ \frac14 (r_+-\xi-r_-+\eta) \right]}\times\\
&\times  I_0\left(\frac12 \sqrt{(r_+-\xi)(\eta-r_-)}\right),
\end{split}
\end{split}
\end{equation}
where $I_0$ is the Bessel function of complex argument
(see, e.g., \cite{WhittakerWatson-1927}).

We shall consider a typical initial distribution where $\rho(x,t=0)$ reaches
an extremum value $\rho_m$ at $x = 0$
and is an even function of $x$: $\rho(-x,t=0) = \rho(x,t=0)$. Such a pulse
has the background density $\rho_0$ and characteristic
width $l\gg\rho_m-\rho_0$ with $u(x,t=0)=0$.
It is convenient to define the inverse functions of the initial $r_\pm$ profiles.
The symmetry of the initial conditions
makes it possible to use the same functions for
$r_+ \in [r_0, r_m]$ and for $r_- \in [-r_m, -r_0 ]$:
\begin{equation} \label{}
\begin{split}
x(r_\pm) =
\begin{cases}
\;\;\,\overline{x}(r_\pm), & \quad \text{if} \quad x >0, \\
-\overline{x}(r_\pm), & \quad \text{if} \quad x < 0.
\end{cases}
\end{split}
\end{equation}
We denote the Riemann invariant at the
background as $r_0 = r_+(x \rightarrow \pm\infty,t=0)$ and
the initial maximum of Riemann invariant as $r_m=r_+(x=0,t=0)$.

Now the curve $\mathcal{C}$ is
represented by the `antidiagonal' $r_- = -r_+=-r$ along which
Eqs.~(\ref{HodTransEqSol}) with $t = 0$ give
\begin{equation}\label{}
\overline{x}(r_-) = \frac{\partial W}{\partial r_-}, \quad \overline{x}(r_+) = \frac{\partial W}{\partial r_+}.
\end{equation}
Hence from
\begin{equation}\label{}
\overline{W} (\xi,\eta) = \int_0^{\xi} \overline{x}(r) dr + \int_0^{\eta} \overline{x}(r) dr
\end{equation}
we get the necessary contraction of $W$ on $\mathcal{C}$. Then $\overline{W}(-r,r)=0$ and
Eqs.~(\ref{m1-284.4}) reduce to
\begin{equation}\label{}
U=\frac{1}{2} \overline{x}(r) R(r,-r;\xi,\eta), \quad V=-\frac{1}{2} \overline{x}(r) R(r,-r;\xi,\eta).
\end{equation}

\begin{figure}[t!]
	\centering
	\includegraphics[width=0.4\textwidth]{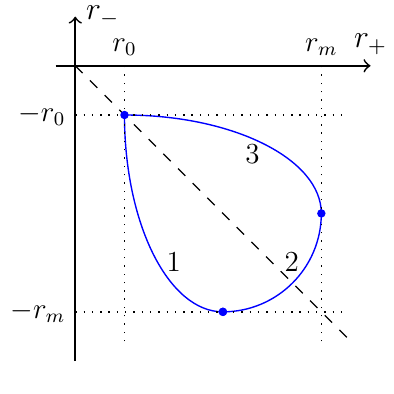}
	\caption{Characteristic plane ($r_-,r_+$). The blue solid curve depicts the
values of the Riemann invariants along the wave at fixed moment of time $t$.}
	\label{fig:one}
\end{figure}

\begin{figure}[t!]
	\centering
	\includegraphics[width=0.5\textwidth]{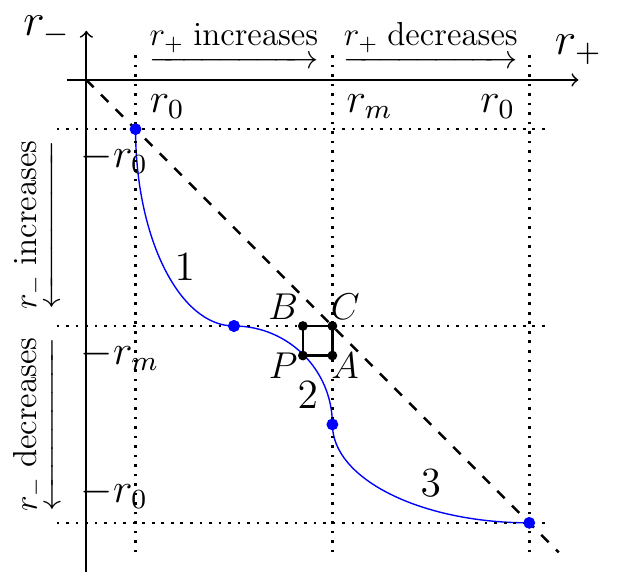}
	\caption{The unfolded hodograph plane ($r_-,r_+$). The blue solid curve
depicts the values of the Riemann invariants along the wave at fixed
moment of time $t$.}
	\label{fig:two}
\end{figure}

It is convenient to divide $x$-axis into several domains
(see \cite{IKP-2019,IKP-2019F,IKP-2020}) which correspond to specific
behavior of the Riemann invariants in each domain.
For a general solution, we numerate these regions by the Arabic
numbers $1$, $2$, and $3$. In region $1$, the Riemann invariant $r_+$
is an increasing function of $x$, and the Riemann invariant $r_-$
is a decreasing function of $x$. In region $2$, both Riemann invariants
increase. In region $3$ invariant $r_+$ decreases, and
$r_-$ increases. Each of these regions requires
its own treatment. The Fig.~\ref{fig:one} shows the hodograph
plane $(r_+,r_-)$-axes. Here, different
regions of the general solution are indicated by various
Arabic numerals. For each such a region, the solution of
the Euler-Poisson equation (\ref{58-3}) $W$ has
its own expression. To find each of these expressions,
we, following Ludford \cite{Ludford-52}, unfold the domain
$[r_0,r_m]\times[-r_m,-r_0]$ into a four times larger region
as is shown in Fig.~\ref{fig:two}.

The potential $W(r_+,r_-)$ takes different
forms in each of the regions labeled $1$, $2$, and $3$ in Fig.~\ref{fig:two} and
can be considered as a single valued function on the unfolded plane. Thus,
from Eq.~(\ref{m1-282.9}) we obtain $W$ in the regions $1$ and $3$
\begin{equation}\label{}
\begin{split}
W^{(1)}(\xi,\eta)=-W^{(3)}(\xi,\eta)=-\int_{-\eta}^{\xi} \overline{x}(r)R(r,-r;\xi,\eta)dr.
\end{split}
\end{equation}
For the region $2$ after integration by parts we obtain
\begin{equation}\label{W2}
\begin{split}
W^{(2)}(\xi,\eta)&=  \left(RW^{(1)}\right)_B+\left(RW^{(3)}\right)_A  \\
& +\int_A^C\left.\left(\frac{\partial R}{\partial r_-}-aR\right)\right|_{r_+=r_m} W^{(3)}dr_-  \\
& -\int_C^B\left.\left(\frac{\partial R}{\partial r_+}-bR\right)\right|_{r_-=-r_m} W^{(1)}dr_+,
\end{split}
\end{equation}
where the coordinates of the relevant points are equal to
$A=(r_m, \eta)$, $B=(\xi, -r_m)$ and $C=(r_m, -r_m)$ (see Fig.~\ref{fig:two}).

In practice, it is convenient to use an approximation suggested in
Refs.~\cite{IKP-2019,IKP-2019F,IKP-2020}. It consists of the assumption
the argument of the Bessel function in Eq.~(\ref{Rim-func}) is small and,
consequently, the Riemann function reduces to the expression
\begin{equation}\label{}
\begin{split}
R(r_+,r_-;\xi,\eta)  &\simeq R(r_+-r_-,  \xi-\eta) \\
&= \exp{\left[ \frac14 (r_+-\xi-r_-+\eta) \right]}.
\end{split}
\end{equation}
In the case of applicability of this approximation, we obtain in regions $1$ and $3$
\begin{equation}\label{W1ex}
\begin{split}
W^{(1)}&(r_-,r_+) =  -W^{(3)}(r_-,r_+) \\
&= \int_{-r_-}^{r_+} \overline{x}(r)R(2r;r_+-r_-)dr, \\
\end{split}
\end{equation}
and in region $2$
\begin{equation}\label{W2ex}
\begin{split}
W&^{(2)}(r_-,r_+)  \\
&= R(r_m,-r_-;r_+-r_-)\int_{-r_-}^{r_m} \overline{x}(r)R(2r,r_m-r_-)dr \\
&+ R(r_++r_m,r_+-r_-)\int_{r_+}^{r_m} \overline{x}(r)R(2r,r_++r_m)dr,
\end{split}
\end{equation}
where we have made
the replacements $\xi \rightarrow r_+$, $\eta \rightarrow r_-$
to return to the notation of Eqs.~(\ref{HodTransEqSol}) and (\ref{58-2}).
When $W$ is known, we can find from (\ref{HodTransEqSol}) the Riemann invariants
as functions of $x$ and $t$, and then the physical variables
are expressed by the formulas
\begin{equation}\label{PhysVar}
\begin{split}
\rho  = \exp{\left[\frac12 (r_+-r_-)\right]}, \qquad
u  = \frac12(r_++r_-).
\end{split}
\end{equation}

At the edges of a finite evolving pulse, the simple-waves regions arise with one
of the Riemann invariants constant. We label such regions by Roman
numerals: the region where the Riemann
invariant $r_-$ decreases and the
Riemann invariant $r_+$ remains constant we denote as $\rom{1}$;
the label $\rom{2}_l$ corresponds to the region
where $r_-$ increases and $r_+$ remains constant;
at last, at the right edge of the momentum, we denote
the regions where $r_-$ remains constant and the Riemann invariant $r_+$ increases
and decreases as $\rom{3}$ and $\rom{2}_r$, respectively.

In the simple-wave regions, where one of the Riemann invariants is constant, the hodograph
transform is not valid. At the initial stages of evolution, when the regions
$\rom{1}$ and $\rom{3}$ do exist, we look for the simple
wave solution in the form
\begin{equation}\label{}
\begin{split}
&x - v_-(r_-,r_0)t  = h(r_-), \qquad \text{for region $\rom{1}$}, \\
&x - v_+(-r_0,r_+)t  = h(r_+), \qquad \text{for region $\rom{3}$},
\end{split}
\end{equation}
where the function $h$ is determined by the boundary
conditions of matching of the intensity
at the boundary between the simple wave
and the general solution (see Eqs.~(\ref{HodTransEqSol})).
Thus we have
\begin{equation}\label{}
\begin{split}
x - v_-(r_-,r_0)t = \frac{\partial W^{(1)}(r_-,r_0)}{\partial r_-},
\end{split}
\end{equation}
for the simple-wave region $\rom{1}$ and
\begin{equation}\label{}
\begin{split}
x - v_+(-r_0,r_+)t = \frac{\partial W^{(3)}(-r_0,r_+)}{\partial r_+},
\end{split}
\end{equation}
for the region $\rom{3}$.

After a certain time of evolution the two simple-wave
regions are formed, which are denoted as $\rom{2}_l$ and $\rom{2}_r$.
Similarly, we can get
\begin{equation}\label{}
\begin{split}
x - v_+(-r_0,r_+)t = \frac{\partial W^{(2)}(-r_0,r_+)}{\partial r_+},
\end{split}
\end{equation}
for the region $\rom{2}_l$ and
\begin{equation}\label{}
\begin{split}
x - v_-(r_-,r_0)t = \frac{\partial W^{(2)}(r_-,r_0)}{\partial r_+},
\end{split}
\end{equation}
for the region $\rom{2}_r$.

Thus we obtain the complete description of
the dispersionless wave evolution.

\begin{figure*}[ht]
	\centering
	\includegraphics[width=1\textwidth]{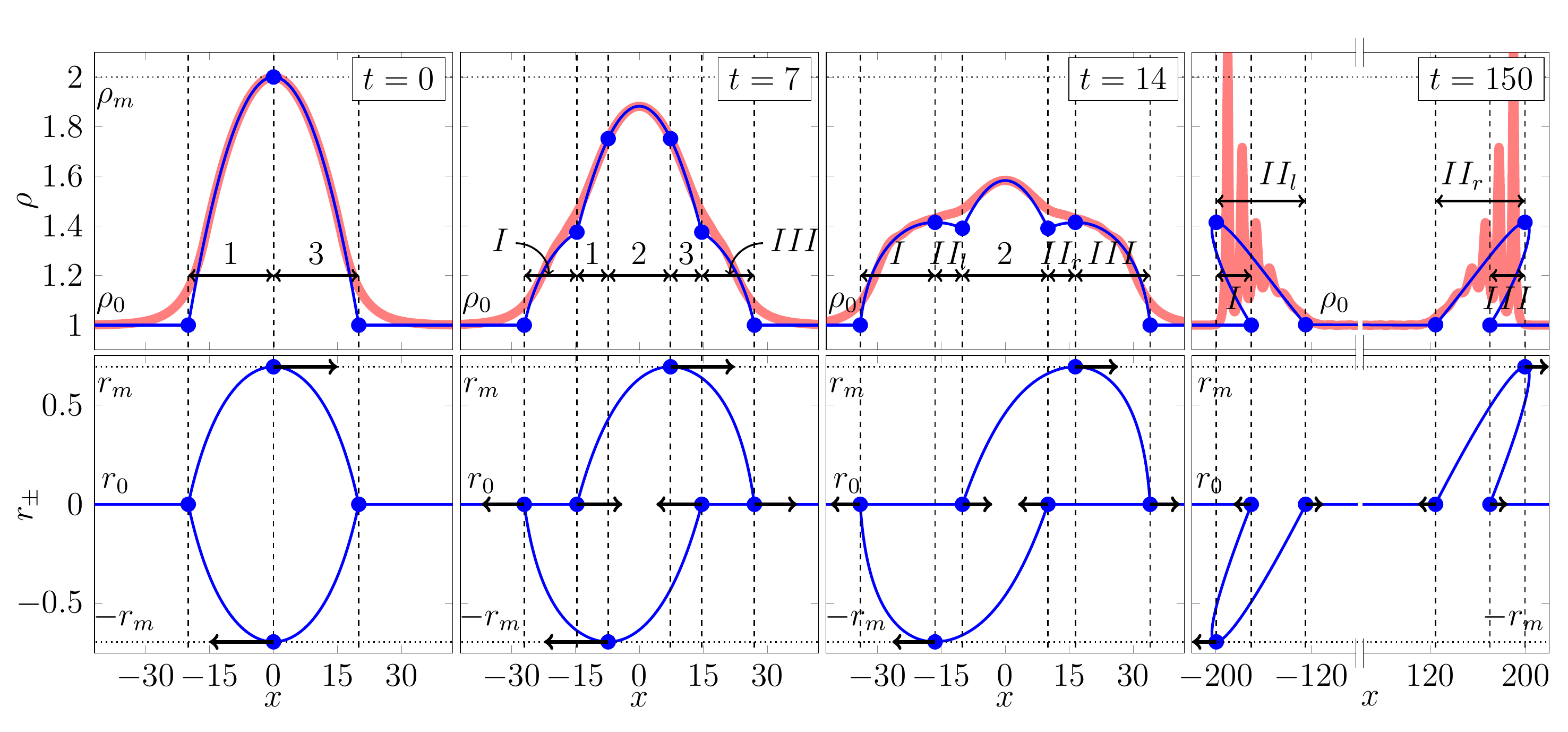}
	\caption{(Top row) Density $\rho$ versus coordinate $x$. A comparison of the analytical solution
(blue thin curve) and direct numerical solution of Eqs.~\ref{PlasmaEq} (red thick curve) for different $t$ is shown.
The initial state has a parabolic shape and it is shown in the left panel. The Arabic numerals
denote the regions of the general solution, and the Roman numerals indicate the regions of simple waves.
(Bottom row) The dependence of the Riemann invariants $r_\pm$ on the spatial coordinate $x$ at
the same moments of time. The right column corresponds to the time after the wave-breaking with formation
of a multi-valued dispersionless solution.}
	\label{fig:three}
\end{figure*}

\begin{figure}[ht]
	\centering
	\includegraphics[width=0.5\textwidth]{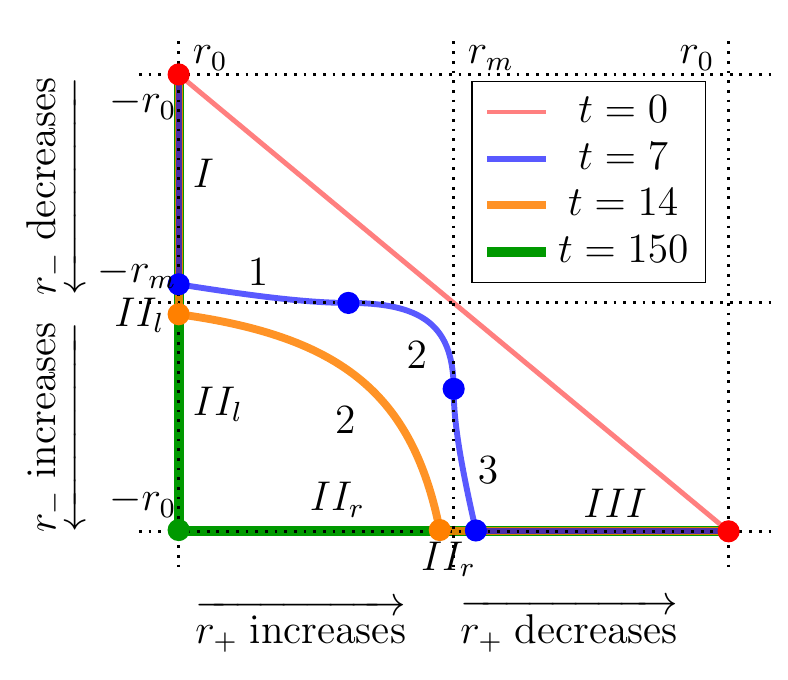}
	\caption{Diagram of the Riemann invariant on the unfolded hodograph plane $(r_+,r_-)$ for the initial
state (\ref{InitialParabola}). Different colors indicate curves at different moments in time.
The region above the red line is unreachable for the initial distribution under consideration.}
	\label{fig:four}
\end{figure}

We shall illustrate the above theory by the example with a parabolic initial distribution of the
density
\begin{equation} \label{InitialParabola}
\begin{split}
\rho(x,t=0) =
\begin{cases}
\rho_0+(\rho_m-\rho_0)(1-\frac{x^2}{l^2}), & \quad |x| \leq l, \\
\rho_0, & \quad |x| > l,
\end{cases}
\end{split}
\end{equation}
where $l$ is a characteristic width of the distribution and the initial flow velocity
is equal to zero, $u(x,t=0)=0$ (see left panel on the top row of Fig.~\ref{fig:three}).
For these conditions the inverse function of $r(x,t=0)$
(see left panel on the bottom row of Fig.~\ref{fig:three})
for the positive branch has the form
\begin{equation}\label{}
\overline{x}(r)=l \sqrt{\frac{\rho_m-e^r}{\rho_m-\rho_0}}.
\end{equation}
As one can see from Fig.~\ref{fig:three}, two simple-wave regions
$\rom{1}$ and $\rom{3}$ appear in the course of evolution,
as well as a region of general solution, which is labeled by
${2}$. Next, regions $1$ and $3$ vanish, and new simple wave
regions $\rom{2}_l$ and $\rom{2}_r$ appear (third column of
Fig.~\ref{fig:three}).  For longer time (right column of Fig.~\ref{fig:three}),
region 2 also vanishes and only simple-wave regions persist:
the initial pulse splits into two simple-wave pulses
propagating in opposite directions.
The corresponding curves on a four-sheeted hodograph plane
are shown in the Fig.~\ref{fig:four}.

Substitution of the initial conditions into our approximate
expressions (\ref{W1ex}) and (\ref{W2ex}) for $W$ yields
\begin{equation} \label{Wexpparb}
\begin{split}
W^{(1)}&(r_-,r_+) =  -W^{(3)}(r_-,r_+) \\
& = - \left(\varPhi(r_+,r_+,r_-)-\varPhi(-r_-,r_+,r_-)\right),\\
W^{(2)}&(r_-,r_+)
 = R(r_m - r_-, r_+ - r_-) \\
 &\times \left(\varPhi(r_m, r_m, r_-)-\varPhi(-r_-, r_m,r_-)\right) \\
 & + R(r_+ + r_m, r_+ - r_-)  \\
 & \times \left(\varPhi(r_m, r_+,-r_m)-\varPhi(r_+, r_+,-r_m)\right),
\end{split}
\end{equation}
where
\begin{equation*} \label{}
\begin{split}
\varPhi(r,\xi,\eta)
&=\frac{l}{\sqrt{\rho_m-\rho_0}} \exp{\left(-\frac{\xi-\eta}{4}\right)} \\
&\times\left( e^{r/2} \sqrt{\rho_m-e^r} + \rho_m \arcsin{
\frac{e^{r/2}}{\sqrt{\rho_m}}} \right).
\end{split}
\end{equation*}

Once $r_+$ and $r_-$
have been determined as functions
of $x$ and $t$ using (\ref{Wexpparb}) and (\ref{tEq}),
the density and velocity profiles are calculated
by means of Eqs.~(\ref{PhysVar}). One obtains a quite exact description of
the initial dispersionless stage of evolution of the pulse, as is
demonstrated in the top row of Fig.~\ref{fig:three}.
At larger time, the density profile at both edges of the
pulse steepens, DSWs are formed and the amplitude of these oscillations
accordingly increases. The dispersionless approximation
subsequently predicts a nonphysical multi-valued profile,
as can be seen in right column of Fig.~\ref{fig:three}.
It is worth noticing that the wave breaking leads to overlapping
of the regions. An example of such
a solution is shown in the right column of Fig.~\ref{fig:four}.
One can see that two simple-wave regions
($\rom{1}$ and $\rom{2}_l$ or $\rom{2}_r$ and $\rom{3}$) overlap.
We shall not consider this phenomenon in detail now,
since such multi-valued regions are nonphysical and,
when dispersion is taken into account, they are replaced by DSWs.

\section{Dispersive shock wave formation} \label{sec4}

We have shown that any localized initial pulse with
initial distributions of $\rho(x, 0)$, $u(x, 0)$ different from some
constant values $\rho_0$, $u_0=0$ on a finite interval of $x$ evolves
eventually into two separate pulses propagating
in opposite directions. These two pulses are called
simple-wave solutions in which one of the
Riemann invariants $r_\pm$ is constant.
Further evolution leads to steepening of the wave
profile and wave breaking followed by formation of a DSW.
Here, in this section, we suppose that the
initial state belongs to such a class of simple waves. To be
definite, we assume that $r_-=u-\ln\rho=-\ln\rho_0=\mathrm{const}$,
that is we consider the right-propagating wave with
\begin{equation} \label{}
\rho=\rho_0 e^u, \quad r_+ = 2u + \ln\rho_0,
\end{equation}
so that the solution of dispersionless equations can be
written as
\begin{equation} \label{dispsol}
x - (u+1)t = \ox(u),
\end{equation}
where $\ox(u)$ is the function inverse to the initial
distribution of the local flow velocity.
Thus, we consider the wave breaking in terms of the flow velocity $u$.

\subsection{Positive Pulse} \label{sec4.1}

\begin{figure}[t]
	\centering
	\includegraphics[width=0.5\textwidth]{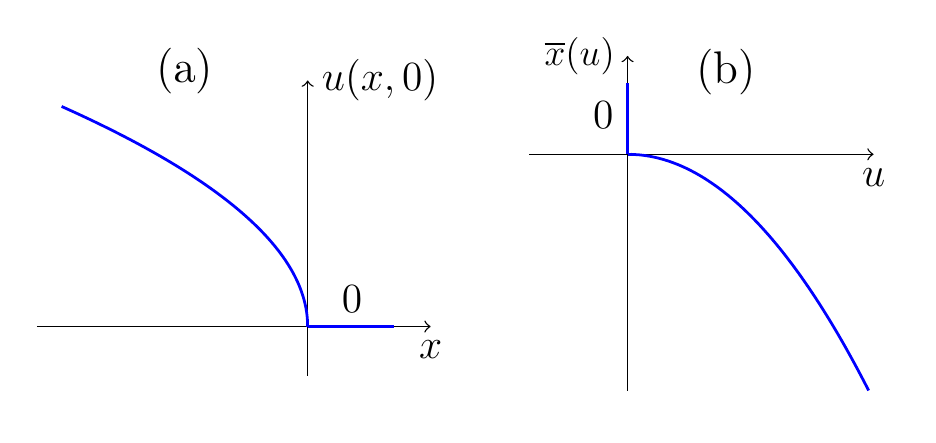}
	\caption{(a) Monotonous positive initial condition and (b) the inverse function $\ox(u)$.}
	\label{fig:five}
\end{figure}

We shall start with a monotonous  pulse with the initial
distribution
\begin{equation}\label{u_init}
u(x,0)=
\begin{cases}
\widetilde{u}(x) & \mbox{if}\quad x<0, \\
0 & \mbox{if}\quad x\ge 0,
\end{cases}
\end{equation}
where $\widetilde{u}(x)$ monotonously increases with decrease of $x$; see Fig.~\ref{fig:five}(a).
We can see at once that if the initial distribution of the flow velocity
$u$ has the form of a hump (``positive pulse''), then the wave
breaking occurs at the front of the pulse with formation of solitons at the front
edge of the DSW and with small-amplitude edge at the boundary with the smooth part
of the pulse described by Eq.~(\ref{dispsol}) (rear edge of the DSW). Therefore,
our task here is to find the law of motion of the small-amplitude edge
by the method of Ref.~\cite{Kamchatnov-19}.

To this end, we have to solve the number of waves conservation law equation~(\ref{eq1})
where $k=k(u)$ is an unknown function to be found. Actually, it was done in \cite{El-05},
and we reproduce here briefly the necessary results in a convenient for us form.
We represent the dispersion relation (\ref{harmlaw}) in the form
\begin{equation}\label{7.1}
\om(k) = k[u+\al(k)],
\end{equation}
that is the function
\begin{equation*}\label{}
\al(k)=\frac{1}{\sqrt{1+k^2/\rho}}, \qquad (0<\al<1),
\end{equation*}
measures deviation of the dispersion law from a linear non-dispersive limit,
and then we have
\begin{equation}\label{7.2}
k(u)=\sqrt{\rho_0e^u\left(\frac{1}{\al^2}-1\right)},
\end{equation}
where $\al=\al(u)$ is an unknown yet function.

With the use of the equation
\begin{equation}\label{7.3}
\frac{\prt u}{\prt t}+(u+1)\frac{\prt u}{\prt x}=0
\end{equation}
corresponding to one of the limiting characteristic
velocities of the Whitham system and equivalent to Eq.~(\ref{rimeq})
for $r_+$, we reduce Eq.~(\ref{eq1}) to the ordinary differential
equation
\begin{equation}\label{7.4}
\frac{d\al}{du}=-\frac{\al(1+\al)^2}{2(1+\al+\al^2)},
\end{equation}
which coincides with the equation obtained in Ref.~\cite{El-05}.
It has to be solved with the boundary condition
\begin{equation}\label{7.7}
\al(0)=1,
\end{equation}
which means that the distance between solitons tends to infinity
at the soliton edge, hence $k(u)\to0$ as
$u\to 0$. Integration of Eq.~(\ref{7.4})
with the boundary condition (\ref{7.7}) is elementary and yields
\begin{equation}\label{7.8}
u(\al)=-2\ln\al-\frac{1-\al}{1+\al}.
\end{equation}

Now, we can find the law
of motion of the small-amplitude edge of the DSW which
propagates with the group velocity equal to, as one can easily obtain,
\begin{equation}\label{7.9}
s_L=\frac{d\om}{dk}=u(\al)+\al^3.
\end{equation}
The small-amplitude edge matches the dispersionless solution (\ref{dispsol}).
Along the path of the small-amplitude edge we have $ dx-s_Ldt = 0$, i.e.,
$x = x_L(u)$ and $t = t(u)$ satisfy the equation
\begin{equation}\label{7.10}
\frac{\prt x}{\prt u}-s_L\frac{\prt t}{\prt u}=0.
\end{equation}
This equation must be compatible with Eq.~(\ref{dispsol}) and this condition yields the equation
\begin{equation}\label{7.11}
(\al^3-1)\frac{dt}{du}-t=\frac{d\ox}{du}
\end{equation}
for the function $t=t(u)$, where $\ox(u)$ is the function inverse to the initial distribution
of $u(x)$; see Fig.~\ref{fig:five}(b). Since we already know the relationship (\ref{7.8}) between $u$ and $\al$,
it is convenient to transform Eq.~(\ref{7.11}) to the independent variable $\al$. To simplify
the notation, we introduce the function
\begin{equation}\label{7.12}
\Phi(\al)=\left.\frac{d\ox}{d\al}\right|_{u=u(\al)}
\end{equation}
and obtain
\begin{equation}\label{7.13}
\frac{\al}{2}(1-\al)(1+\al)^2\frac{dt}{d\al}-t=\Phi(\al).
\end{equation}
Its solution with the initial condition $t(0) = 0$ reads
\begin{equation}\label{7.14}
\begin{split}
t(\al) &=2\exp{\left[\frac{1}{1+\al}\right]} \frac{\al^2}{(1-\al)^{1/2}(1+\al)^{3/2}} \\
 & \times \int_{1}^{\al} \frac{\Phi(z)\exp{(-1/(1+z))}}{z^3\sqrt{1-z^2}} dz
\end{split}
\end{equation}
and, consequently,
\begin{equation}\label{7.15}
x_L(\al)=[u(\al)+1]t(\al)+\ox[u(\al)].
\end{equation}
The formulas (\ref{7.14}) and (\ref{7.15}) define in a parametric form the law of
motion $x=x_L(t)$ of the small-amplitude edge.

\begin{figure}[t]
	\centering
	\includegraphics[width=0.5\textwidth]{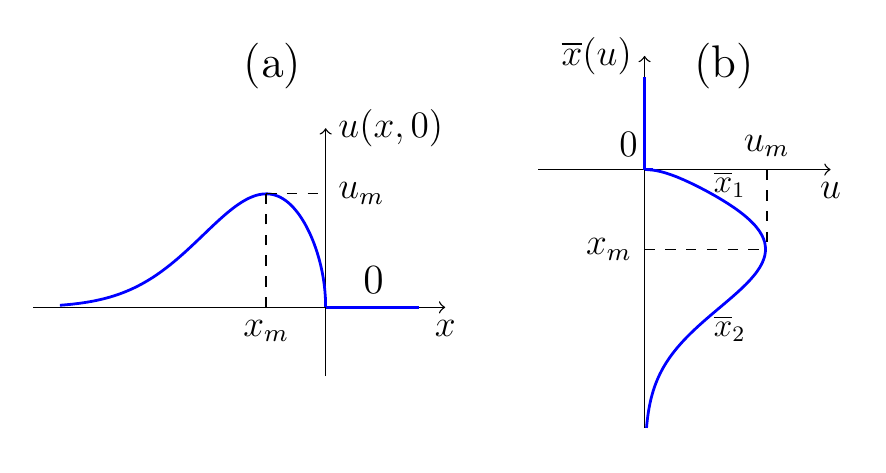}
	\caption{(a) Localized positive initial pulse;
(b) the inverse function $\ox(u)$ is represented by two branches $\ox_1(u)$ and $\ox_2(u)$.}
	\label{fig:six}
\end{figure}

In the case of a localized initial pulse shown in Fig.~\ref{fig:six}
the inverse function becomes two-valued and we denote its
two branches as $\ox_1(u)$ and $\ox_2(u)$.
The initial distribution $u(x)$
has a single maximum $u_m$ at $x=x_m$ and $u(x)\to 0$ fast enough as $x\to-\infty$
and $u(x)\to 0$ with vertical tangent line at $x\to-0$.
The above formulas obtained for monotonous initial distribution are applicable as long as
the small-amplitude edge propagates along the branch $\ox_1(u)$, so that the solution can
be derived by replacement of the function $\Phi(\al)$ by $\Phi_1(\al)=\left. d\ox_1/du\right|_{u=u(\al)}$.
When the small-amplitude edge reaches the maximum $u_m$ at the moment $t_m$,
then the equation (\ref{7.13}) with
$\Phi(\al)$ replaced by $\Phi_2(\al)=\left. d\ox_2/du\right|_{u=u(\al)}$ should be solved with
the boundary condition
$t(\al_m)=t_m,$
where $\al_m$ is the root of the equation $u(\al_m)=u_m$.
Hence, for $t>t_m$ the motion of the small-amplitude is determined by the formula
\begin{equation}\label{8.18}
\begin{split}
t(\al) &=2\exp{\left[\frac{1}{1+\al}\right]} \frac{\al^2}{(1-\al)^{1/2}(1+\al)^{3/2}} \\
& \times \left(\int_{1}^{\al_m} \frac{\Phi_1(z)\exp{(-1/(1+z))}}{z^3\sqrt{1-z^2}} dz + \right.\\
& \qquad \left. \int_{\al_m}^{\al} \frac{\Phi_2(z)\exp{(-1/(1+z))}}{z^3\sqrt{1-z^2}} dz\right).
\end{split}
\end{equation}

The law of motion of the soliton edge of DSW cannot be found by this method, since the
relation (\ref{eq2}) does not hold during evolution of the pulse. However, as was remarked
in Ref.~\cite{Kamchatnov-19}, this relationship can be used in vicinity of the moment when
the small-amplitude edge reaches the point with $u=u_m$.
We write the soliton dispersion law (\ref{sollaw}) in the form
\begin{equation}\label{9.1}
\tom(\tk)=\tk[u+u\tal(\tk)],
\end{equation}
that is
\begin{equation*}\label{}
\tal(\tk)=\frac{1}{\sqrt{1-\tk^2/\rho}},\qquad (\tal>1),
\end{equation*}
Substitution of Eq.~(\ref{9.1}) together with
\begin{equation}\label{9.3}
\tk(c)=\sqrt{\rho_0e^u\left(1-\frac{1}{\al^2}\right)}   .
\end{equation}
into Eq.~(\ref{eq2}) gives
\begin{equation}\label{9.4}
\frac{d\tal}{du}=-\frac{\tal(1+\tal)}{2(1+\tal+\tal^2)},
\end{equation}
which, according El's approach, should be
solved with the boundary condition
\begin{equation}\label{9.5}
\tal(u_m)=1,
\end{equation}
which means that the solitons widths become infinitely large ($\tk\to0$) and their
amplitude is infinitely small at the small-amplitude edge. As
a result, we get
\begin{equation}\label{9.6}
u(\tal)=u_m-2\ln\tal - \frac{1-\tal}{1+\tal}
\end{equation}
and
\begin{equation}\label{10.1}
s_R=\left.\frac{\tom}{\tk}\right|_{u=0}=\tal(0).
\end{equation}
Here $u<u_m$, $\tal$ increases with decrease of $u$ and reaches its maximal value at
$u=0$.
This value of the leading soliton velocity in DSW is reached at large enough time when
solitons near the leading edge are well separated from each other.

\begin{figure}[t]
	\centering
	\includegraphics[width=0.45\textwidth]{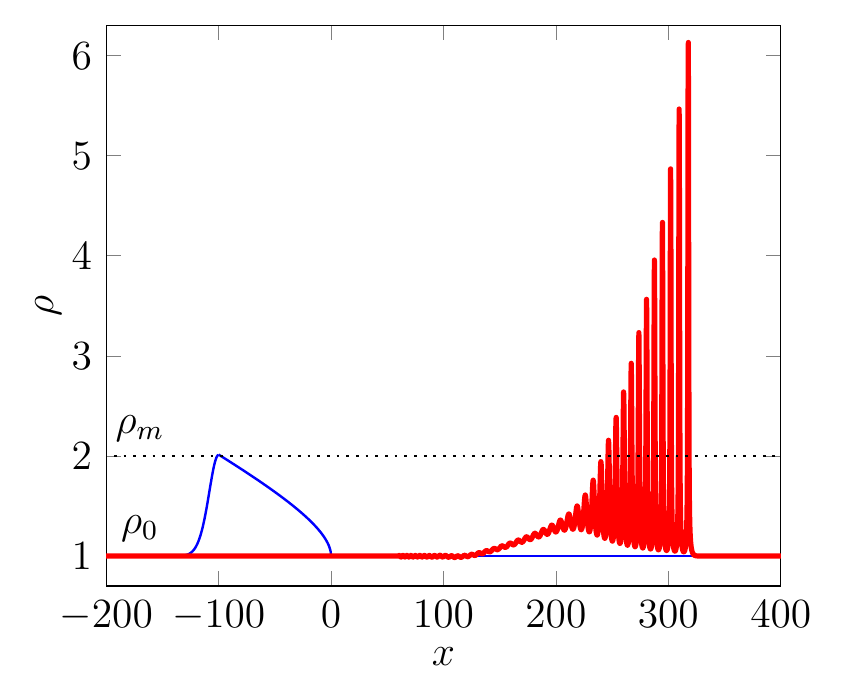}
	\caption{Simple-wave initial state (\ref{InitialStatePositive}) with $\rho_0=1$,
		$x_0=-100$ and $B=0.07$ is shown by the blue (thin) solid curve. Density profile at $t=250$ is shown by the red (thick) solid curve).}
	\label{fig:seven}
\end{figure}

\begin{figure}[t]
	\centering
	\includegraphics[width=0.45\textwidth]{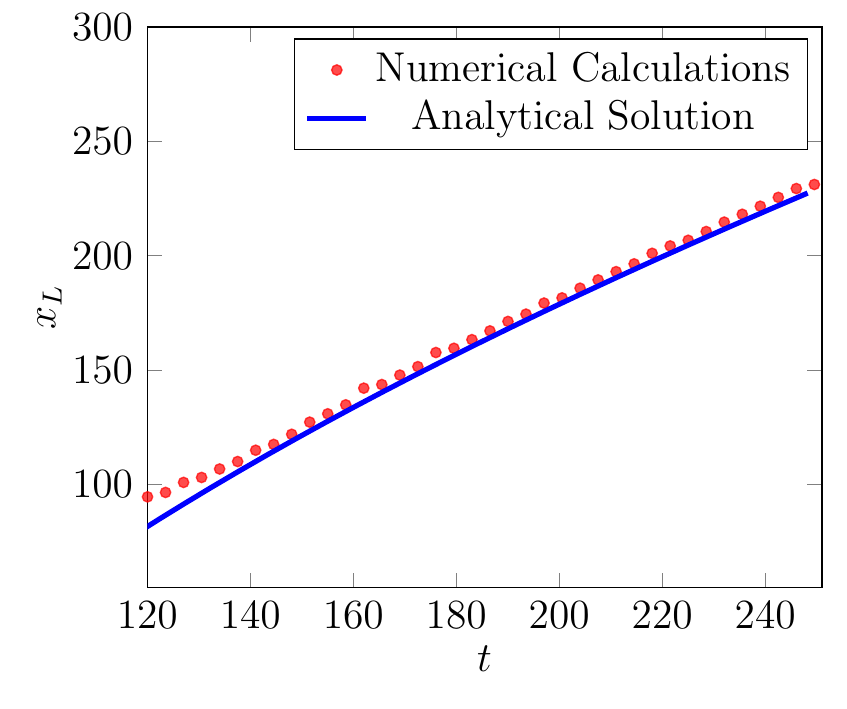}
	\caption{Comparison of the numerical solution of Eqs.~(\ref{PlasmaEq}) (red dots) and the analytical parametric solution (\ref{7.15}) and (\ref{8.18}) (blue (solid) curve) for the small-amplitude edge of DSW. The initial state parameters are indicated in the caption Fig.~\ref{fig:seven}.}
	\label{fig:eight}
\end{figure}

\begin{figure}[t]
	\centering
	\includegraphics[width=0.45\textwidth]{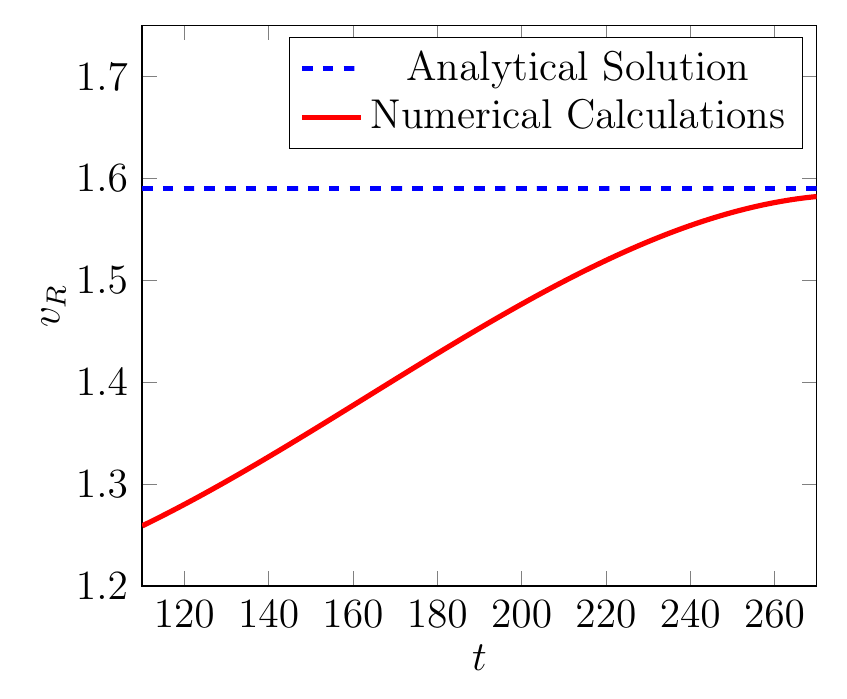}
	\caption{Velocity of the soliton edge as a function of time (red (solid) line). Blue (dashed) line indicates the asymptotic value of the velocity of this edge given by Eq.~(\ref{10.1}).}
	\label{fig:nine}
\end{figure}

We have compared our analytical approach with numerical simulations of the Eqs.~(\ref{PlasmaEq}).
We have chosen the initial condition in the form
\begin{equation} \label{InitialStatePositive}
\begin{split}
u(x,t=0) =
\begin{cases}
A \exp{\left(-\frac{(x - x_0)^2}{4\delta(x_0-\delta)}\right)}, & \quad \text{if} \quad x \le \delta, \\
B\sqrt{-x}, & \quad \text{if} \quad x > \delta,
\end{cases}
\end{split}
\end{equation}
where
$$
A = {B} \;\exp{\left(-\frac{1}{4}\left(1-\frac{x_0}{\delta}\right)\right)} {\sqrt{-\delta}}\;, \quad \delta = x_0+0.5.
$$
So that density distribution has the form (see blue (thin) curve in Fig.~\ref{fig:seven})
\begin{equation} \label{initialrho}
\begin{split}
\rho = \rho_0 e^{u(x,t=0)}.
\end{split}
\end{equation}
This initial state is split into two branches as shown in the Fig.~\ref{fig:six}.
The typical form of the DSW
generated by such a pulse is represented in the same figure by a red (thick) curve.
Eqs.~(\ref{7.15}) and (\ref{8.18}) give the parametrical dependence of the small-amplitude edge of the DSW $x = x_L(t)$ shown in Fig.~\ref{fig:eight} by a blue solid curve. As we can see, it agrees reasonably well with the result of numerical solution shown by a red dots. We determine the position of small-amplitude edge numerically by means of an approximate extrapolation of the envelopes of the wave at this edge. For the right solitonic edge of the DSW, we can find the asymptotic propagation velocity, which for a given initial state is approximately equal to $s_R=1.59$ and is shown in Fig.~\ref{fig:nine} by a dashed line. Fig.~\ref{fig:nine} shows that the numerically determined velocity of a given edge tends to the asymptotic analytical value $s_R$.

\subsection{Number of solitons}

For asymptotically large time $t\to\infty$, a localized pulse shown in Fig.~\ref{fig:six}(a)
transforms into a train of bright solitons and we have found the velocity (\ref{10.1})
of the leading soliton. Another important characteristic of this evolution is the number
$N$ of solitons produced from a pulse with a given initial distribution $u_0(x)=u(x,0)$.
To calculate this number $N$, we turn to an important remark made by Gurevich and
Pitaevskii in \cite{gp-87} that the number of waves entering per unit of time into
the DSW region at its small-amplitude edge is given by
\begin{equation}\label{eq10-k}
  \frac{dN}{dt}=\frac1{2\pi}\left(k\frac{\prt\om}{\prt k}-\om\right),
\end{equation}
where the right-hand side is calculated at the values of the small-amplitude edge parameters
at the moment of time $t$. According to Eq.~(\ref{eq1}), the frequency $\om(k)/(2\pi)$ plays the
role of the flux of the number of waves, so the right-hand side of Eq.~(\ref{eq10-k}) can
be interpreted as a flux calculated with account of Doppler shift due to motion of the
edge with the group velocity $d\om/dk$.

If we consider a localized initial pulse, then all waves inside the DSW transform eventually into
solitons and their number is given by the integral \cite{kamch-20b}
\begin{equation}\label{k2}
  N=\frac1{2\pi}\int_0^{\infty}\left(k\frac{\prt\om}{\prt k}-\om\right)dt.
\end{equation}
Since the function $k(\al)$ is defined by Eqs.~(\ref{7.2}), (\ref{7.8}), and $t(\al)$
is given by (\ref{7.14}), (\ref{8.18}), this integral can be calculated directly. 
However, it is instructive to transform it much simpler form. To this end we
substitute the above formulas into (\ref{k2}) to get
\begin{equation}\label{k3}
  \begin{split}
  & N=\frac1{\pi}\int_{\al_m}^1\exp\left(-\frac{1-\al}{2(1+\al)}\right)\frac{\sqrt{1-\al^2}}{\al^2(1+\al)}\times\\
  & \times\Bigg\{2\exp\left(\frac1{1+\al}\right)\frac{\al^2}{(1-\al)^{1/2}(1+\al)^{3/2}}\times\\
  &\times\int_{\al_m}^{\al}\frac{(\Phi_2-\Phi_1)e^{-\frac1{1+z}}dz}{z^3\sqrt{1-z^2}}+\Phi_2(\al)-\Phi_1(\al)\Bigg\}d\al.
  \end{split}
\end{equation}
We notice that the double integral here can be reduced to the ordinary one by integration by
parts and after collecting all terms and some simplifications we obtain
\begin{equation}\label{k4}
   N=\frac1{\pi}\int_{\al_m}^1e^{-\frac{1-\al}{2(1+\al)}}(\Phi_2-\Phi_1)
   \frac{1+\al+\al^2}{\al^3(1+\al)^2}d\al.
\end{equation}
Now, taking into account the definition (\ref{7.12}) of $\Phi$, Eq.~(\ref{7.4}), and the initial condition
$\ox(\al)=-2\ln\al-\frac{1-\al}{1+\al}$ obtained from (\ref{7.8}) at $t=0$, we arrive at the
final result
\begin{equation}\label{k5}
  N=\frac1{2\pi}\int_{-\infty}^0k[u_0(x)]dx,
\end{equation}
where $k(u)$ is defined in Eq.~(\ref{7.2}). This formula has a general nature. It was suggested in
Refs.~\cite{egkkk-07,egs-08} as a consequence of continuation of solution of the Whitham
modulation equations to the dispersionless region. For some other equations it was justified
by a similar calculations in Ref.~\cite{kamch-20b}.

\subsection{Negative pulse}  \label{sec4.2}

\begin{figure}[t]
	\centering
	\includegraphics[width=0.5\textwidth]{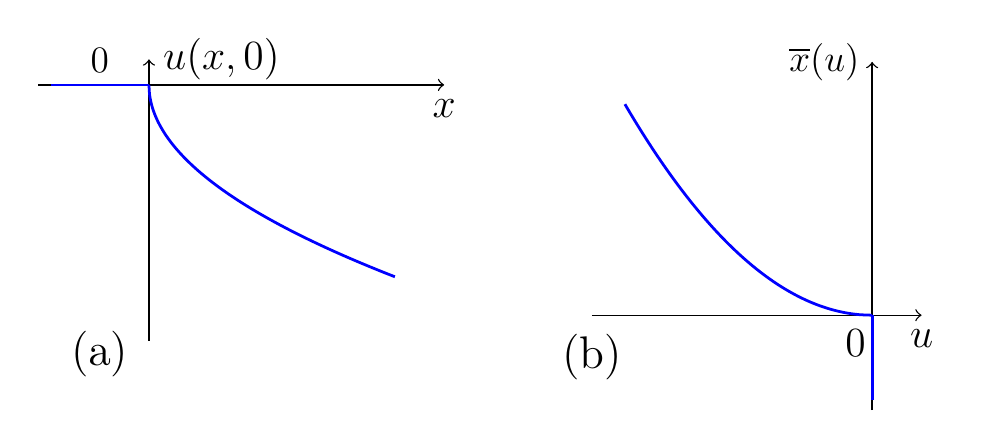}
	\caption{(a) Monotonous negative initial condition; (b) the inverse function $\ox(u)$.}
	\label{fig:ten}
\end{figure}

\begin{figure}[t]
	\centering
	\includegraphics[width=0.5\textwidth]{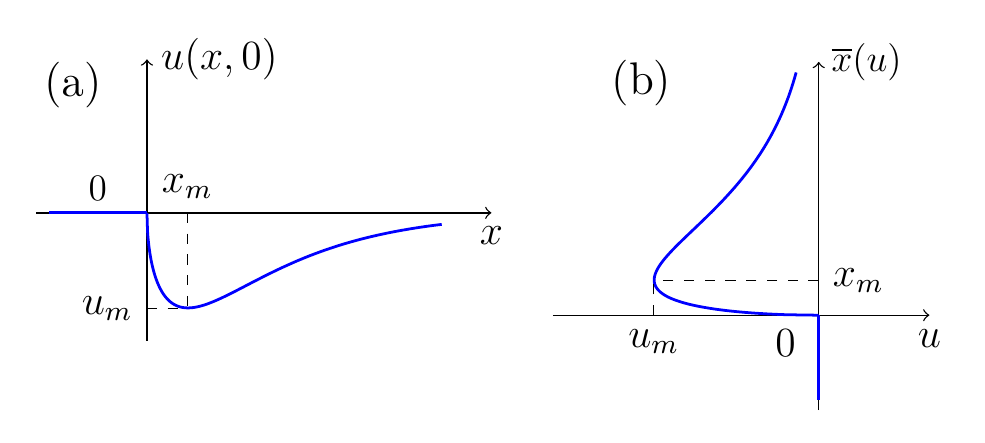}
	\caption{(a) Localized negative initial pulse;
(b) the inverse function $\ox(u)$ is represented by two branches $\ox_1(u)$ and $\ox_2(u)$.}
	\label{fig:eleven}
\end{figure}

Now we turn to the situation sketched in
Figs.~\ref{fig:ten}(a) with a monotonous pulse
in the initial distribution of the velocity
\begin{equation}\label{8.23}
u(x,0)=
\begin{cases}
0 & \mbox{if}\quad x<0, \\
-\widetilde{u}(x) & \mbox{if}\quad x\ge 0,
\end{cases}
\end{equation}
The inverse function $\ox(u)$ is shown in Fig.~\ref{fig:ten}(b).
As was indicated
in Ref.~\cite{Kamchatnov-19}, in situations of this
kind Eq.~(\ref{dispsol}) is fulfilled at the soliton edge
located at the boundary with the smooth part of the
pulse which is described in dispersionless
approximation by Eq.~(\ref{dispsol}). The small-amplitude
edge of the DSW is located at the
boundary with the quiescent medium where $u=u_0=0$.
Eq.~(\ref{eq2}) can be transformed in the same way as it was done
in Eqs.~(\ref{9.1})-(\ref{9.4}),
but now Eq.~(\ref{9.4}) must be solved with the boundary condition
\begin{equation}\label{8.24}
\tal(0)=1.
\end{equation}
This gives
\begin{equation}\label{8.25}
u(\tal)=-2\ln\tal-\frac{1-\tal}{1+\tal},
\end{equation}
where $\tal>1$,
and consequently the trailing soliton velocity is equal to
\begin{equation}\label{10.2}
s_R=\frac{\tom}{\tk}=u(\tal)+\tal.
\end{equation}
It corresponds to the characteristic velocity of the limiting
Whitham equation
\begin{equation}\label{11.1}
\frac{\prt x}{\prt u}-[u(\tal)+\tal]\frac{\prt t}{\prt u}=0.
\end{equation}
The condition that this equation must be compatible with the solution (\ref{dispsol}) of the
dispersionless approximation at the soliton edge of the DSW yields the differential equation
for the function $t=t(u)$ at this edge,
\begin{equation}\label{11.2}
(\tal-1)\frac{dt}{du}-t=\frac{d\ox}{du}.
\end{equation}
Introducing the function $\Phi(\tal)=\left.d\ox/du\right|_{u=u(\tal)}$ we cast it to the form
\begin{equation}\label{12.1}
\frac{\tal(\tal-1)(\tal+1)^2}{2(1+\tal+\tal^2)}\frac{dt}{d\tal}+t=-\Phi(\tal)
\end{equation}
with $\tal$ as an independent variable. Since we assume that the initial profile breaks at the
moment $t=0$ at the rear edge where $u=u_0$, {\it i.e.,} $\tal=1$, this equation should be solved with
the boundary condition
\begin{equation}\label{12.2}
t(1)=0.
\end{equation}
As a result, we obtain
\begin{equation}\label{12.3}
\begin{split}
t(\tal)&=\frac{2\tal^2}{(\tal-1)^{3/2}(\tal+1)^{1/2}}\exp{\left[\frac{1}{\tal+1}\right]} \times \\
\times & \int_1^{\tal}\Phi(z)\frac{(z-1)^{1/2}(1+z+z^2)}{z^3(z+1)^{3/2}}\exp{\left[-\frac{1}{z+1}\right]}dz.
\end{split}
\end{equation}
The coordinate of the soliton edge is given by
\begin{equation}\label{13.1}
x_R(\tal)=[u(\tal)+1]t(\tal)+\ox[u(\tal)],
\end{equation}
so that Eqs.~(\ref{12.3}), (\ref{13.1}) define the law of motion $x=x_R(t)$ of this edge
in parametric form.
Generalization of this theory on localized pulses (see Fig.~\ref{fig:eleven}) is straightforward.

At asymptotically large time $t\to\infty$ we have $\tal\to1$ and
$t(\tal)\approx\mathcal{A}(\tal-1)^{-3/2}$, where
\begin{equation}\label{14.2}
\mathcal{A}=-
\int{\Phi(z)\frac{(z-1)^{1/2}(1+z+z^2)}{z^3(z+1)^{3/2}}\exp{\left[-\frac{1}{z+1}\right]}dz},
\end{equation}
where the integral should be taken over the whole initial pulse
($\int\Phi=\int_1^{\tal_m}\Phi_1+\int_{\tal_m}^1\Phi_2$). In this limit
$u(\tal)\approx u_m-3/2\;(\tal-1)$ and simple calculation yields
\begin{equation}\label{14.3}
x_R(t)\approx (u_m-1)t-\frac{3}{2}\mathcal{A}^{2/3}t^{1/3}.
\end{equation}

As was shown in Ref.~\cite{Kamchatnov-19}, we can
find velocity of the small-amplitude edge
at asymptotically large time for a localized initial pulse by solving
Eq.~(\ref{eq1}) with (\ref{harmlaw}). Then this equation reduces again to Eq.~(\ref{7.4}),
but now it should be solved with the boundary condition
\begin{equation}\label{15.1}
\al(u_m)=1
\end{equation}
where $u_m$ is the maximal value of the velocity in the
initial distribution $\widetilde{u}(x)$.
The wave number given by Eq.~(\ref{7.2}) tends here to zero. This gives
\begin{equation}\label{15.2}
u=u_m-2\ln{\alpha}+\frac{\al-1}{\al+1}.
\end{equation}
At the small-amplitude edge we have $u=0$
and the group velocity (\ref{7.9}) plays the role
of the left-edge velocity,
\begin{equation}\label{15.3}
\frac{1-s_L^{1/3}}{1+s_L^{1/3}}+\frac23 \ln{s_L} =u_m.
\end{equation}
Solution of this equation gives asymptotic velocity of small-amplitude edge.

\begin{figure}[t]
	\centering
	\includegraphics[width=0.45\textwidth]{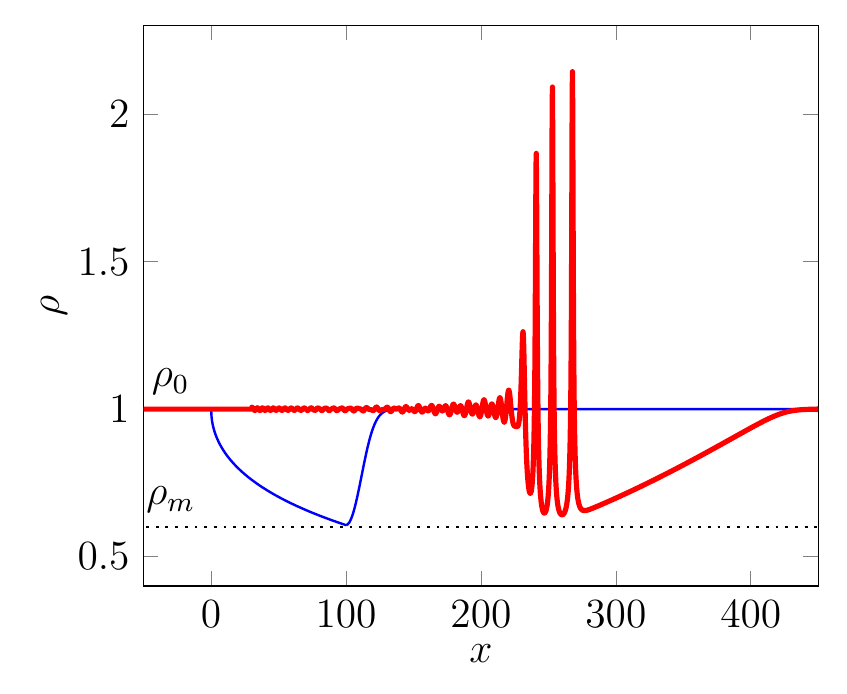}
	\caption{Simple-wave initial state (\ref{InitialStateNegative}) with $\rho_0=1$,
		$x_0=100$ and $B=0.05$ is shown by the blue (thin) solid curve. Density profile at $t=300$ is shown by the red (thick) solid curve).}
	\label{fig:twelve}
\end{figure}

\begin{figure}[t]
	\centering
	\includegraphics[width=0.45\textwidth]{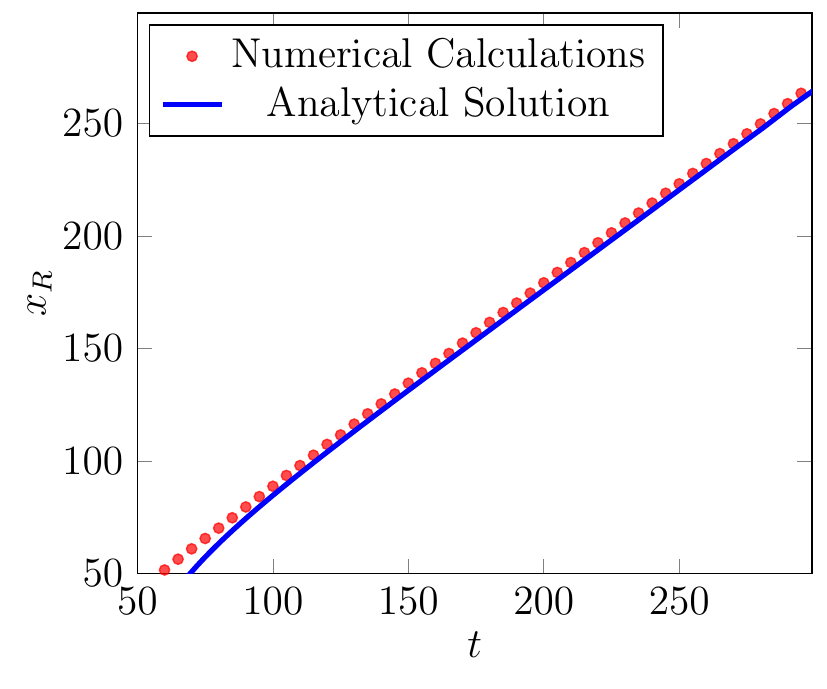}
	\caption{Comparison of the numerical solution of Eqs.~(\ref{PlasmaEq}) (red dots) and the analytical parametric solution (blue (solid) curve) for the soliton edge of DSW. The initial state parameters are indicated in the caption Fig.~\ref{fig:twelve}.}
	\label{fig:thirteen}
\end{figure}

To compare analytical results with numerical calculations, we took the initial state
\begin{equation} \label{InitialStateNegative}
\begin{split}
u(x,t=0) =
\begin{cases}
-A \exp{\left(-\frac{(x - x_0)^2}{4\delta(x_0-\delta)}\right)}, & \quad \text{if} \quad x \ge \delta, \\
-B \sqrt{x}, & \quad \text{if} \quad x < \delta,
\end{cases}
\end{split}
\end{equation}
where
$$
A = {B} \;\exp{\left(-\frac{1}{4}\left(1-\frac{x_0}{\delta}\right)\right)} {\sqrt{\delta}}\;, \quad \delta = x_0-0.5.
$$
So that density distribution has the form that can be obtained from Eq.~(\ref{initialrho}) (see blue (thin) curve in Fig.~\ref{fig:twelve}).
The red (thick) curve in Fig.~\ref{fig:twelve} shows the density distribution at $t=300$.
Propagation path of the soliton edge is depicted in Fig.~\ref{fig:thirteen}.
Analytical results are shown by a blue (solid) line.
As we can see, it agrees reasonably well without any fitting
parameter with the result of numerical solution shown by
a red dotted line. 

The above comparison of the analytical predictions
with the numerical solution demonstrates quite convincingly that the method suggested here gives accurate
enough description of pulses whose evolution obeys nonintegrable equations.

\section{Conclusion} \label{sec5}

In this paper, we have studied dynamics of ion-sound pulses in plasma
for arbitrary form of the initial distribution of simple wave type.
This investigation generalizes earlier theories
\cite{gke-90-plasma,El-05} applied to the step-like initial conditions
only and corresponds to realistic experimental situations \cite{ABS-68,TBI-70}.
Besides that, we derived an asymptotic formula for the number of solitons
generated from a localized positive initial pulse.
The analytical results obtained here are confirmed by numerical simulations.
All that demonstrates that the method of Ref.~\cite{Kamchatnov-19} is quite
effective and can be used for prediction of typical parameters of DSWs
generated in real experiments.

\begin{acknowledgments}
	The reported study was funded by RFBR, Project No. 19-32-90011.
\end{acknowledgments}

\end{document}